\documentclass{article}%
\usepackage{eurosym}
\usepackage[top=3cm, bottom=3cm, left=2.5cm, right=2.5cm]{geometry}
\usepackage{amsmath}
\usepackage{amsfonts}
\usepackage{amssymb}
\usepackage{graphicx}%
\usepackage{epsfig}
\usepackage{float}
\begin{document}

\title{\textbf{Mass Spectra and Semileptonic Decays of  Doubly Heavy $ \Xi $ and $ \Omega $  Baryons}}

\author{Zahra Ghalenovi\footnote{$z_{-}ghalenovi@kub.ac.ir$}  $ ^{1} $  , Cheng-Ping~Shen$ ^{2} $  and\\
 Masoumeh Moazzen Sorkhi$ ^{1} $\\
$ ^{1} $Department of Physics, Kosar University of Bojnord, Bojnourd 94156-15458, Iran\\
$ ^{2} $Key Laboratory of Nuclear Physics and Ion-beam Application (MOE) and \\
Institute of Modern Physics, Fudan University, Shanghai 200443, China
}

\maketitle

\begin{abstract}
In the framework of a non-relativistic quark model, the mass spectra of
the ground and excited states of doubly heavy $\Xi$ and $\Omega$ baryons are calculated.
 We estimate the mass difference between the $ \Omega $ and
corresponding $ \Xi $ baryons as $M_{\Omega}-M_{\Xi}\simeq178 $ MeV
for all the states containing $ cc,~bc $, or $ bb $  quarks.
  A simple form of the universal Isgur-Wise function, as the transition
form factor between the doubly heavy baryons, is introduced.  Working in the
close-to-zero recoil limit, we investigate the $ b \rightarrow c $
semileptonic decay widths and branching fractions of the doubly heavy baryons.
The obtained results are compared with other theoretical predictions.
\end{abstract}



\section{Introduction}

The investigation of doubly heavy baryons is of great interest
in understanding quantum chromodynamics (QCD) inspired potential model, the non-relativistic QCD factorization theory, etc., at the hadronic scale. During the
last few years many theoretical progresses have been achieved in doubly
heavy baryon spectroscopy~\cite{Weng2018,Garcilazo2016,
Shaha2017,Brown2014,Aliev12013}. In 2002,
the lightest doubly charmed baryon $ \Xi ^{+}_{cc}$
was observed with a statistical significance of 6.3$\sigma$
with a measured mass of $(3519 \pm 1)$ MeV in the decay mode $ \Xi ^{+}_{cc}\to \Lambda_c^+ K^- \pi^+$ by the SELEX collaboration~\cite{Mattson2002}.
This state was subsequently confirmed by the same collaboration
in another decay mode $ \Xi ^{+}_{cc} \to p D^+ K^-$~\cite{Ocherashvili2005}.
However, negative results in searching for the $ \Xi ^{+}_{cc}$ were reported by FOCUS~\cite{nfocus}, BaBar~\cite{nbabar}, or Belle~\cite{nbelle}
collaborations. In 2017, the doubly charmed baryon $ \Xi ^{++}_{cc}$
was first observed by the LHCb collaboration via the decay mode $ \Xi ^{++}_{cc} \to  \Lambda_c^+ K^-\pi^+\pi^+ $
with a measured mass $ (3621.40\pm0.72\pm0.14\pm0.27)$ MeV~\cite{LHCb2017},
where the uncertainties are statistical, systematic, and from
the limited knowledge of the $\Lambda_c^+$ mass, respectively,
and confirmed in another decay mode $ \Xi ^{++}_{cc} \to   \Xi ^{+}_{c} \pi^+$~\cite{LHCb2018}.
The measured mass is about 100 MeV higher than that of $ \Xi ^{+}_{cc}$ determined by the SELEX collaboration~\cite{Mattson2002}.
The lifetime of the $ \Xi_c^{++}$ was measured to be $ \tau( \Xi_{cc}^{++})=(0.256^{+0.024}_{-0.022}\pm 0.014)$ ps~\cite{LHCb20188}.
The $ \Xi ^{++}_{cc}$ observation demonstrates how really the LHC is a powerful discovery machine,
stimulating the theoretical studies of mass spectra of doubly heavy baryons.
Therefore, the predictions for doubly heavy baryon
masses have become a subject of renewed interest.

Other than the double-charm $ \Xi_{cc}$  baryons, the beauty-charm  and also double-beauty baryons are the
different kinds of doubly heavy baryons, which have not been found yet.
Recently, several experimental efforts have been made on the exclusive channels
$ \Xi^0_{bc}\rightarrow D^0 p K^-$~\cite{lhcb1} and $ \Xi^0_{bc}\rightarrow
\Xi_c^+ \pi^-$~\cite{lhcb2}  to search for the beauty-charm baryons, but no signals were
observed.
The $ \Xi_{bc}$ baryons are expected to have smaller sizes than  the
$\Xi_{cc}$  states. In comparison to the $\Xi_{cc}$, searching for the
$ \Xi_{bc}$  are more complicated and fewer states could be produced at
the LHC. To overcome this difficulty, an inclusive decay channel
$ \Xi_{bc}\rightarrow \Xi_{cc}^{++}+X$ has been proposed~\cite{qqin}, where $ X $
stands for all the possible particles.

For the singly charmed baryons, some semileptonic decays have been calculated
in theory~\cite{Li2021,Hsiao2020,Lu2021,Faustov2019,
Faustov2016,Gutsche2016} and measured in experiments~\cite{Ablikim2017,Ablikim2018,
Li2021a,Li2021b,Ammar2002}. However, for the doubly heavy baryons, no
experimental data on semileptonic decays are reported and only a limited
number of theoretical calculations are available.
As there are more doubly heavy baryons that may be discovered in
the future, proposing a theoretical model for their structures
 is essential. The investigations of the inclusive and exclusive semileptonic
decays of doubly heavy  baryons are of special interest for two basic
 reasons: they play an important role in the calculation of fundamental
parameters of the electroweak standard model and provide a
useful tool to extract the Cabibbo-Kobayashi-Maskawa (CKM) matrix element $ V_{cb}$.
Some characteristics of semileptonic decays, such as the transition form factors
and exclusive decay rates, also provide information about the internal
structures of heavy baryons and the strong and weak interactions inside them.

Some theoretical calculations to the semileptonic weak decays of doubly heavy baryons have been provided  based on the QCD sum rules (QCDSR)~\cite{Kiselev2001,Shi2020b},
covariant confined quark model~\cite{Gutsche2019}, Bethe-Salpeter  equation~\cite{Yu2019},
heavy quark spin symmetry (HQSS)~\cite{Hernandez2008}, relativistic quark model (RQM)~\cite{Lyubovitskij2003,Lyubovitskij2001},
non-relativistic quark model (NRQM)~\cite{Albertus2008},
 or heavy diquark effective theory (EFT)~\cite{qqin,Shi2020a}.
However, the calculations of semileptonic decays presented by different approaches
lead to essentially different values for the decay widths.

In this paper, we present a description of the properties of heavy baryons containing two heavy quarks (charm or beauty quark)
within the framework of  non-relativistic quark model proposed in Ref.~\cite{ghalenovi2018}. Firstly, we calculate
 the mass spectra of the ground and orbitally excited states of doubly heavy
 $ \Xi $ and $ \Omega $ baryons, including $ \Xi_{cc},~\Xi_{bc},~\Xi_{bb},~\Omega_{cc},~\Omega_{bc} $, and $ \Omega_{bb} $ states, in the hypercentral constituent
  quark model. Then, we focus on the studies of $ b\rightarrow c $ semileptonic decays of the ground states of doubly
  heavy  $ \Xi $ and $ \Omega $ baryons. We proceed our model close to the
 zero recoil point in which the form factors of these transitions can be
 expressed by a few universal functions~\cite{Faessler2009}. Considering the doubly baryon states can be produced sizably at LHC, especially LHC has started Run3 data taking, our results can be tested in the near future.

This paper is structured as follows. In Section~\ref{mass}  we present our predictions for the ground and also orbitally excited states of the doubly heavy baryons using the
 results obtained in our previous work. In Section~\ref{semi decay}
 we introduce a universal function as the transition form factor
 to study the semileptonic decays of doubly charm and bottom heavy
 baryons close to the zero recoil point and present our numerical
 results for semileptonic decay widths and branching fractions of
 specific modes. Section~\ref{Summary} includes a short summary.

\section{Doubly heavy baryon spectra} \label{mass}
We start with a brief review of our previous work~\cite{ghalenovi2018} in which the mass spectra and
radiative transitions of  $ \Sigma_b $ and $ \Lambda_b $ baryons are studied within the hypercentral
constituent quark model~\cite{Isgur1978,Capstick1986,Ferretti2011,Ghalenovi:2012iu,
Lonsdale:2017mzg,Menapara:2021dzi}. The hypercentral quark
model contains a few free parameters, whose values are
obtained by fitting the baryon spectrum. Once the parameters are obtained,
the model is completely determined and we can make our
predictions for the baryon properties. In Ref.~\cite{ghalenovi2018}
 we introduced a phenomenological potential model, and solved the 
baryonic three-body equation in a non-relativistic limit by choosing the Killingbeck potential
$ V(x)=ax^{2}+bx-\frac{c}{x} $. The potential
parameters  $a,~b$, and $c$ are constant.\\
 We apply the Ansatz method~\cite{ansatz1,Ghalenovi:2014swa,Ghalenovi:2017xxv} 
to solve the Schr\"{o}dinger equation. In the hypercentral approach, 
the hyperradial part of the wave function is determined by
\begin{equation}\label{Schro 2}
[\frac{d^2}{dx^2}+\frac{5}{x}\frac{d}{dx}-\frac{\gamma(\gamma+4)}{x^2}]\psi_\gamma(x)=-2m[E_\gamma-V(x)]\psi_\gamma(x).
\end{equation}

Assuming the transformation $\psi(x)=x^{-5/2}\phi(x)$ we get
\begin{equation}\label{Schro 3}
\phi_{\gamma}^{''}(x)+[\varepsilon-a_1x^{2}-b_1x+\frac{c_1}{x}-\frac{(2\gamma+3)(2\gamma+5)}{4x^2}]\phi_{\gamma}(x)=0.
\end{equation}
We make use of the following Ansatz 

\begin{equation}  \label{ansatz1}
{\phi_{\gamma}(x)=exp[-\frac{1}{2} \alpha x^2-\beta x+\delta lnx],}
\end{equation}

where $ \alpha,\beta $ and $ \delta $  parameters are determined 
in terms of the potential parameters. By substitution of equation {\ref{ansatz1}}
into equation {\ref{Schro 3}} and comparing the coefficient of {$x$}
on both sides of the new equation we can evaluate
the energy eigenvalues and normalized eigenfunctions for the baryon 
states as follows (for details see Refs.~\cite{ghalenovi2018,Ghalenovi:2017xxv}):
\begin{equation}\label{E}
E_{\gamma}=(2\gamma+6)\frac{w}{2}-\frac{2mc^2}{(2\gamma+5)^2}
\end{equation}
and
\begin{equation}\label{psi22}
\psi_{\gamma}(x)=N_{\gamma}x^{-\frac{5}{2}}\phi_{\gamma}(x)=N_{\gamma}x^{\gamma}exp(-\frac{m w}{2}x^2-\frac{2mc}{(2\gamma+5)}x),
\end{equation}
where $ x $ is the hyperradius and $ w=\sqrt{\frac{2a}{m}} $ is the oscillating frequency.
  $ m $, $\gamma $ and $N_{\gamma}$  are the  reduced mass, angular quantum number
and normalization constant, respectively. Baryon mass is
obtained by summing the quark masses, energy eigenvalues,
and  hyperfine interaction
potential treated as a perturbation:
\begin{equation}\label{Mass}
M_{\rm baryon}=\sum_{i=1}^{3}m_i+E_{\gamma}+<H_{S}>.
\end{equation}
Here, $<H_{S}>$ is the expectation value of the hyperfine
 spin-spin interactions given as~\cite{ghalenovi2018,
Ghalenovi:2014swa}
\begin{equation}   \label{Hs}
H_S=\Sigma_{i<j}A_{S}(\frac{1}{\sqrt{\pi}\sigma_{S}})^3exp(\frac{-x^2}{\sigma_S})(\overrightarrow{s_{i}}.\overrightarrow{s_j}),
\end{equation}
where  $\overrightarrow{s_i}$ is the spin operator of the
 $i^{\rm th}$ quark, and $\sigma_S$ and $ A_S $ are constant.
The isospin values of the strange and heavy quarks are zero
and therefore, in the case of doubly heavy baryons the
isospin dependent terms are not included in the hyperfine
interactions.  The spin-orbit interaction has values smaller than $0.01$ GeV in the hyperfine
contributions and therefore, we neglect it in our
calculations. All of the model parameters listed in Table~\ref{tab:parameters} are taken from
our previous work~\cite{ghalenovi2018}. We take the
experimentally measured mass of the $ \Xi_{cc}^{++}$~\cite{LHCb2018}  to determine the
mass of the charm quark. Our calculated results of the masses of the ground and
excited baryon states including $P$-wave, $D$-wave, and $F$-wave  are listed in Tables~\ref{tab:mass1}
 and~\ref{tab:mass2},  and compared with the ones
obtained from different models of RQM,
NRQM, QCDSR, HQSS, and EFT~\cite{Ebert2004,Albertus2010,Azizi2013,
Roberts2008,Yoshida2015,Shah2017,Soto2021,Shah2016}.
Table~\ref{tab:mass1} shows that our
results for the masses of the $ \Xi_{cc},~\Xi_{bc},~\Omega_{cc} $,
  and $ \Omega_{bc} $ are in good agreement with those
reported in Ref.~\cite{Ebert2004}. For the masses of the
$ \Xi_{bb}$ and $\Omega_{bb} $ states, our predictions are
close to the results from Refs.~\cite{Azizi2013,Roberts2008,
Yoshida2015}.

 \begin{table}[htbp]
 \caption{Quark-model parameters, where $ q $ refers to the light quarks.}
 \label{tab:parameters}
 \begin{center}
\begin{tabular}{cccc} \hline
Parameter~~&Value&Parameter~~&Value\\ \hline
$m_{q}$  & 320 MeV&$w$  & 1.557 $ {\rm fm}^{-1}$\\
$m_{s}$  & 440 MeV&$b$ & 5.79 $ {\rm fm}^{-2}$ \\
$m_{c}$  & 1360 MeV&$A_s$&67.4 ${\rm fm}^2$\\
$m_{b}$  & 4670 MeV &$\sigma_S$& 4.76 fm\\
\hline

\end{tabular}
  \end{center}
   \end{table}

\begin{table}[H]
\caption{Masses of the ground states of doubly heavy baryons
(in GeV). The sign $``^*"$ refers to the $ s=\frac{3}{2} $ baryons.}
\label{tab:mass1}
 \centering
   \begin{center}
{\begin{tabular}{cccccccc} \hline
Baryon~~& Content~~&Our results~~& RQM~\cite{Ebert2004}~~&HQSS~\cite{Albertus2010}~~ &QCDSR~\cite{Azizi2013}~~&NRQM~\cite{Roberts2008}~~& NRQM~\cite{Yoshida2015}~~\\
\hline
$ \Xi_{cc} $&$ qcc $&  3.620& 3.620&3.613& &3.676&3.685\\
$ \Xi_{cc}^* $&$ qcc $&  3.653&3.727&3.707&3.690&3.753&3.754 \\
$ \Xi_{bc} $&$ qbc $&  6.958&6.933&6.928& &7.020& \\
$ \Xi_{bc}^* $  &$ qbc $&  6.991&6.980&6.996& 7.250&7.078& \\
$ \Xi_{bb} $  &$ qbb $& 10.322&10.202&10.198& &10.340 &10.314\\
$ \Xi_{bb}^* $  &$ qbb $& 10.355& 10.237&10.237&10.400&10.367&10.339 \\
$ \Omega_{cc} $&$ scc $& 3.798 & 3.778&3.712&&3.815&3.832 \\
$ \Omega_{cc}^* $&$ scc $& 3.831 & 3.872&3.795&3.780&3.876&3.883\\
$ \Omega_{bc} $&$ sbc $& 7.137  & 7.088&7.013& &7.147&\\
$ \Omega_{bc}^* $  &$ sbc $& 7.170 & 7.130&7.075&7.300&7.191&\\
$ \Omega_{bb} $  &$ sbb $& 10.500 & 10.359&10.269&&10.456&10.447\\
$ \Omega_{bb}^* $  &$ sbb $& 10.533  & 10.389&10.307&10.500&10.486&10.467\\
\hline
\end{tabular}}
 \end{center}
\end{table}

\begin{table}[H]
\caption{Masses of the orbitally excited states of  doubly heavy baryons (in GeV).}
\label{tab:mass2}
 \centering
 \begin{center}
{\begin{tabular}{ccccc cccc} \hline
State&Baryon& Our results &NRQM\cite{Shah2017}&EFT\cite{Soto2021} &Baryon& Our results&NRQM\cite{Shah2016}&NRQM\cite{Yoshida2015} \\
\hline
&$ \Xi_{cc} $  & 3.928 &  3.853&4.028  &$ \Omega_{cc} $  &4.106&3.964 & 4.086 \\
&$ \Xi_{cc}^* $  & 3.961 &3.862&  4.079 &$ \Omega_{cc}^* $  &4.139&3.972 &4.199 \\
$P$-wave&$ \Xi_{bc} $  & 7.266 &  7.140&  &$ \Omega_{bc} $  & 7.4457& 7.375& \\
&$ \Xi_{bc}^* $  & 7.299 & 7.157&    &$ \Omega_{bc}^* $  &7.478&7.381 &  \\
&$ \Xi_{bb} $  & 10.631 & 10.502& 10.386  &$ \Omega_{bb} $  &10.808&10.634 &10.607 \\
&$ \Xi_{bb}^* $  & 10.664 & 10.510& 10.404  &$ \Omega_{bb}^* $  &10.841&10.636 & 10.796\\
\hline
&$ \Xi_{cc} $  & 4.237 & 4.026  &  4.321  &$ \Omega_{cc} $  & 4.414& 4.133&4.263 \\
&$ \Xi_{cc}^* $  & 4.269 & 4.035& 4.376  &$ \Omega_{cc}^* $  &4.474&4.141 & 4.265\\
$D$-wave&$ \Xi_{bc} $  & 7.575 & 7.307&   &$ \Omega_{bc} $  & 7.754&7.807& \\
&$ \Xi_{bc}^* $  & 7.607 & 7.312&   &$ \Omega_{bc}^* $  &7.786&7.812 &  \\
&$ \Xi_{bb} $  & 10.939 & 10.658& 10.585 &$ \Omega_{bb} $  &11.1177&10.783 &10.723 \\
&$ \Xi_{bb}^* $  & 10.972 &  10.660& 10.610 &$ \Omega_{bb}^* $  &11.150& 10.785 & 10.730\\
\hline
&$ \Xi_{cc} $  & 4.545 & 4.185& 4.569 &$ \Omega_{cc} $  &4.723&4.287 &4.555 \\
&$ \Xi_{cc}^* $  &4.577  & 4.216& 4.644 &$ \Omega_{cc}^* $  &4.755&4.313 &4.600 \\
$F$-wave&$ \Xi_{bc} $  & 7.883 & 7.451&  &$ \Omega_{bc} $  &8.062&7.702 & \\
&$ \Xi_{bc}^* $  & 7.916 & 7.468&  &$ \Omega_{bc}^* $  & 8.095&7.721&  \\
&$ \Xi_{bb} $  & 11.274 & 10.797& 10.757 &$ \Omega_{bb} $  & 11.425& 11.136& 11.028\\
&$ \Xi_{bb}^* $  & 11.280 & 10.805& 10.792  &$ \Omega_{bb}^* $  &11.458&11.142 &11.059 \\
\hline
\end{tabular}}
 \end{center}
\end{table}

Using the masses from our calculations shown in Table~\ref{tab:mass1},
we can get the mass differences $ \Delta_M $ between
$ \Omega $ and corresponding $ \Xi $ doubly heavy baryons:
\begin{eqnarray}
M_{\Omega_{bb}}-M_{\Xi_{bb}}=178 \quad {\rm MeV}, \\
M_{\Omega_{cc}}-M_{\Xi_{cc}}= 178 \quad {\rm MeV}, \\
M_{\Omega_{bc}}-M_{\Xi_{bc}}=179 \quad {\rm MeV}.
\end{eqnarray}
We obtain $ \Delta_M\sim178 $ MeV for all of the
$ M_{\Omega}-M_{\Xi} $ splittings, compared with $ \Delta_M=100\pm10$ MeV
reported in  Ref.~\cite{Kiselev2001} and $ \Delta_M=155\sim158 $  MeV
predicted in Ref.~\cite{Ebert2004}. In other works referenced
in Table~\ref{tab:mass1}, $ \Delta_M$ has different values
for the different $ M_{\Omega}-M_{\Xi} $ splittings.

\section{Semileptonic decays of doubly  heavy baryons} \label{semi decay}
To study the semileptonic transitions of baryons we
 need the form factors, which can be
parameterized in different approaches. Some earlier works~\cite{Faessler2009,Georgi1990, Carone1991,Flynn2008}
simplified the transition form factors using
different methods. The authors of Ref.~\cite{Faessler2009}
studied the form factors and semileptonic decays of
doubly heavy baryons using a relativistic covariant
 quark model (RCQM).
 According to Refs.~\cite{Faessler2009,Isgur1991,Ebert20066},
in the heavy quark limit, the expressions
 for the decay rates can be simplified and the weak
 transition form factors between doubly heavy baryons
 can be expressed through the single Isgur-Wise (IW)
 function $\eta(\omega)$
\begin{eqnarray} \label{eta}
F_1(\omega)=G_1(\omega)=\eta(\omega), \quad \quad \quad \quad \quad\quad\\
F_2(\omega)=F_3(\omega)=G_2(\omega)=G_3(\omega)=0 \nonumber.
\end{eqnarray}

\begin{figure}[htbp]
\centering
\includegraphics[width=0.4\textwidth]{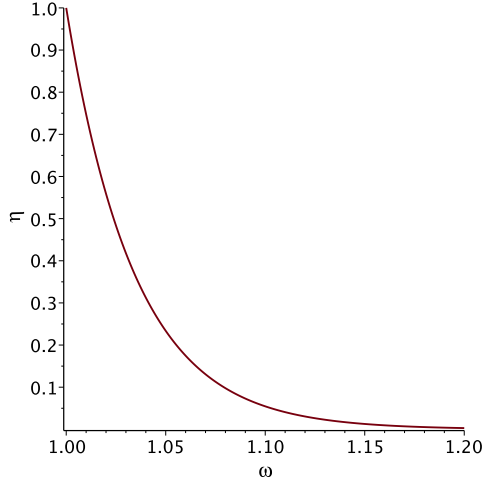}
\caption{ Variation of the universal $\eta(\omega)$ function versus $\omega$ for  $bb\rightarrow bc$ semileptonic transitions with $\Lambda_B=3$ GeV. }
\label{fig:1}
\end{figure}

\noindent In Ref.{~\cite{Faessler2009}} a closed-form expression
 has been derived for the IW function
 {$\eta(\omega)$} using a Gaussian Ansatz for the three-quark 
correlation function in the heavy quark 
limit and close to zero recoil point.
In the present work we take the universal function  {$\eta(\omega)$}
 calculated by Ref.{~\cite{Faessler2009}} which depends on 
the kinematical parameter $ \omega $ and is given by
\begin{equation} \label{eta}
\eta(\omega)=exp(-3(\omega-1)\frac{m_{cc}^2}{\Lambda_B^2})
\end{equation}
with $m_{cc}=2m_c$ for the $ bc\rightarrow cc $  weak
transitions~\cite{Faessler2009}. The parameter $ \Lambda_B $   characterizes
the size of the given baryon and represents the extension
of the distribution of the constituent quarks in the baryon.
For the doubly heavy baryons the size of parameter
$ \Lambda_B $ is allowed to vary in the range  $ 2.5 \leq
 \Lambda_B \leq 3.5$ GeV~\cite{Faessler2009}. The values of
parameter $ \Lambda_B $  are fixed using data on
fundamental properties of mesons and baryons such as leptonic
 decay constants, magnetic moments, and radii.
The dependence of the universal $\eta(\omega)$
 function on  $ \omega $ with $\Lambda_B=3$ for the
$bb \rightarrow bc$ transitions is shown in Fig.~\ref{fig:1}.
For the slope $ \rho^2 $ of the $ \eta(\omega)=1-\rho^2
(\omega-1)+...$, one obtains
\begin{equation}
\rho^2=-\frac{d\eta(\omega)}{d\omega}\vert_{\omega=1}=3\frac{m_{cc}^2}{\Lambda_B^2}.
\end{equation}

By replacing $ m_{cc}$ with $m_{bb} $ in the IW function
 one can obtain the results for the
$ bb\rightarrow bc $ transitions. Accordingly,
the slope of the IW function for the
$ bb\rightarrow bc $ transitions is obtained
from Eq.~(\ref{eta}) by replacing
$ m_{cc}$ with $m_{bb} $ if one uses the
same size of parameter $ \Lambda_B $ in both
cases~\cite{Faessler2009}. Close to zero recoil, the
IW functions for $ bb\rightarrow bc $ and
$ bc\rightarrow cc $ transitions explicitly contain
the flavor factors $ m_{cc}$ and $ m_{bb} $, and there exists only spin symmetry.
There is no dependence on the light quark masses.
At zero recoil ($ \omega=1 $) there exists a
spin-flavor symmetry and $ \eta(1)=1 $ means that
$ bb\rightarrow bc $ and $ bc\rightarrow cc $
transitions are identical.

According to Ref.~\cite{Faessler2009}, in zero recoil limit,
the expressions for the semileptonic decay widths can be
simplified considerably and we can get the decay rates using
the IW function $ \eta(\omega) $ and the following relations:
\begin{equation} \label{gamma11}
\Gamma_{\frac{1}{2}\rightarrow\frac{1}{2}}=\frac{G_F^2\vert V_{bc}\vert ^2M_{i}^5r^4}{12\pi^3}\int_1^{\omega_{\rm max}}d\omega\sqrt{\omega^2-1}(l^+_1(\omega)\eta^2(\omega)+l^-_1(\omega)\eta^2(\omega)),
\end{equation}
\begin{equation}
\Gamma_{\frac{1}{2}\rightarrow\frac{3}{2}}=\frac{G_F^2\vert V_{bc}\vert ^2M_{i}^5r^4}{12\pi^3}\int_1^{\omega_{\rm max}}d\omega\sqrt{\omega^2-1}l_2(\omega)\eta^2(\omega),
\end{equation}
\begin{equation}
\Gamma_{\frac{3}{2}\rightarrow\frac{1}{2}}=\frac{G_F^2\vert V_{bc}\vert ^2M_{i}^5r^4}{24\pi^3}\int_1^{\omega_{\rm max}}d\omega\sqrt{\omega^2-1}l_3(\omega)\eta^2(\omega),
\end{equation}
\begin{equation}
\Gamma_{\frac{3}{2}\rightarrow\frac{3}{2}}=\frac{G_F^2\vert V_{bc}\vert ^2M_{i}^5r^4}{24\pi^3}\int_1^{\omega_{\rm max}}d\omega\sqrt{\omega^2-1}(l^+_4(\omega)\eta^2(\omega)+l^-_4(\omega)\eta^2(\omega)),
\end{equation}
where  $ r=M_2/M_1$,~$\omega_{\rm max}=(1+r^2)/2r $,
\begin{equation}
l^\pm_1(\omega)=(\omega\mp1)(3\omega_{\rm max}\pm1-2\omega),
\end{equation}
\begin{equation}
l_2(\omega)=2(\omega+1)(\omega_{\rm max}-\omega+\frac{\omega^2-1}{6r}),
\end{equation}
\begin{equation}
l_3(\omega)=2(\omega+1)(\omega_{\rm max}-\omega+\frac{(\omega^2-1)r}{6}),
\end{equation}
and
\begin{equation} \label{l4}
l^\pm_4(\omega)=\frac{4}{9}(l^\pm_1(\omega)\frac{3+2\omega^2}{2}\pm(\omega_{max}\pm1)(\omega^2-1)).
\end{equation}
The parameter $G_F $ is the Fermi Coupling constant, $ V_{bc} $ is
the CKM matrix element and its value is
 $ V_{bc}=0.04 $, $ M_1 $ and  $ M_2 $ are the initial
 and final state baryon masses, respectively. We take
the following values of $ \omega_{\rm max} $ for different transitions:
\begin{eqnarray}
\omega_{\rm max}[\Xi_{bb}\rightarrow \Xi_{bc}]=1.07, \quad \omega_{\rm max}[\Omega_{bb}\rightarrow \Omega_{bc}]=1.07,\\
\omega_{\rm max}[\Xi_{bc}\rightarrow \Xi_{cc}]=1.22, \quad \omega_{\rm max}[\Omega_{bc}\rightarrow \Omega_{cc}]=1.20.
\end{eqnarray}

\begin{figure}[H]
\centering
\includegraphics[width=0.4\textwidth]{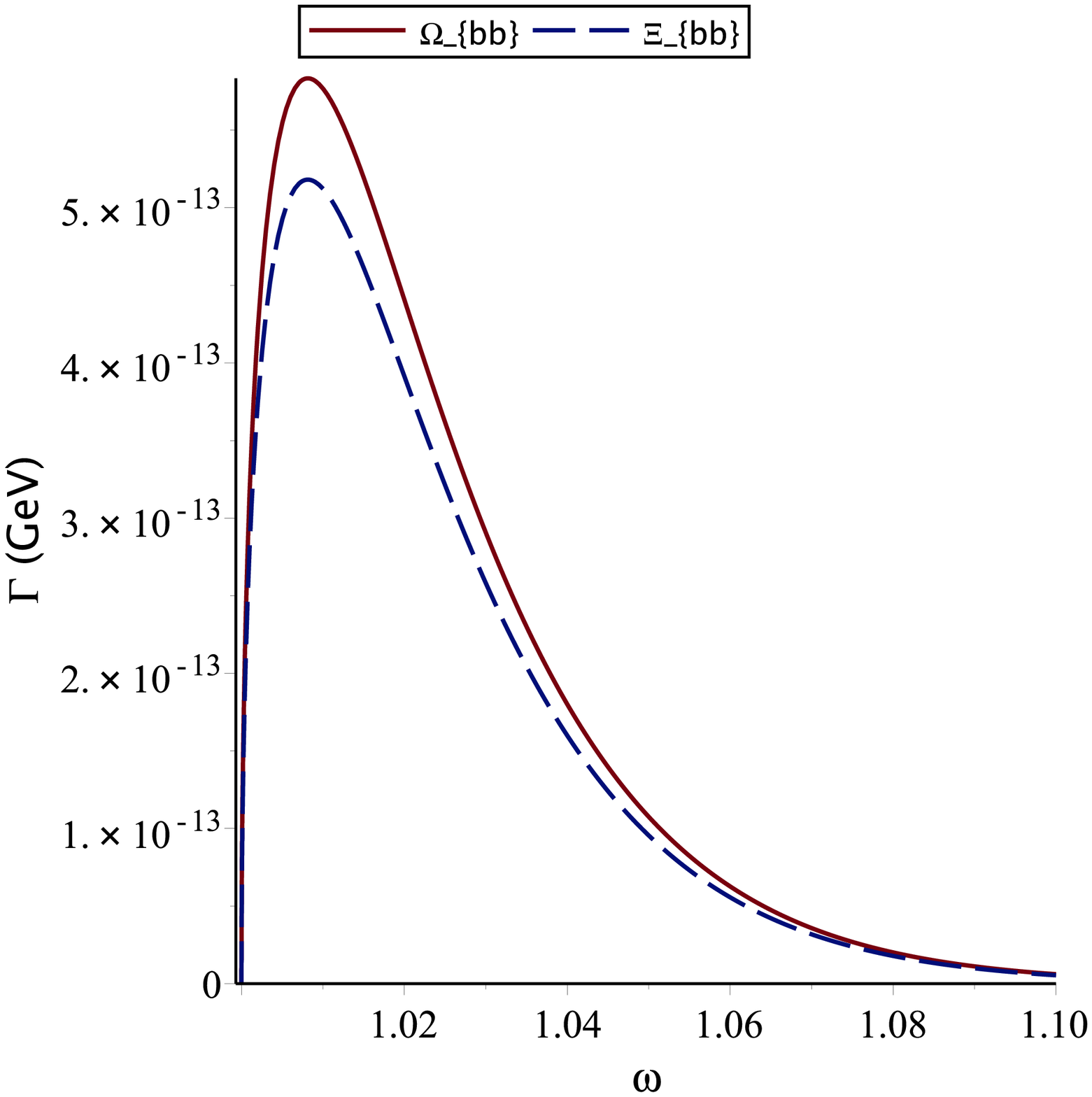}

\caption{ Behavior of semileptonic decay widths versus $ \omega $ for  $\Omega_{bb}\rightarrow \Omega_{bc}\ell \bar {\nu}_{\ell}$ and  $\Xi_{bb}\rightarrow \Xi_{bc}\ell \bar {\nu}_{\ell}$ ($\ell=e$ or $\mu$) transitions ($ \Lambda_B=3 $ GeV). }
\label{fig:2}
\end{figure}

\begin{figure}[H]
\centering
\includegraphics[width=0.4\textwidth]{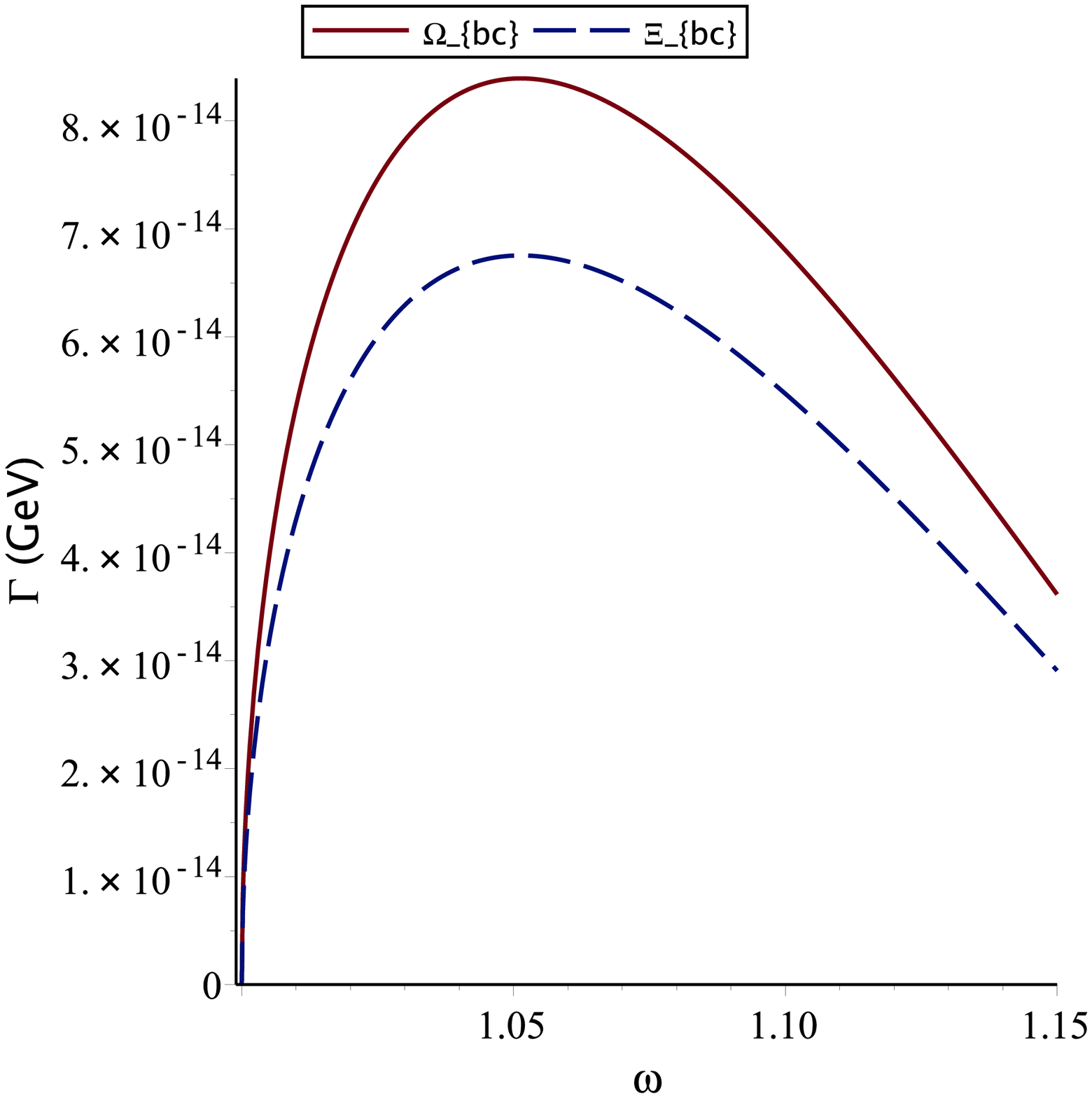}
\caption{ Behavior of semileptonic decay width versus $ \omega $ for  $\Omega_{bc}\rightarrow \Omega_{cc}\ell \bar {\nu}_{\ell}$ and $\Xi_{bc}\rightarrow \Xi_{cc}\ell \bar {\nu}_{\ell}$ ($\ell=e$ or $\mu$) transitions ($ \Lambda_B=3 $ $GeV$). }
\label{fig:3}
\end{figure}

\begin{figure}[H]
\centering
\includegraphics[width=0.4\textwidth]{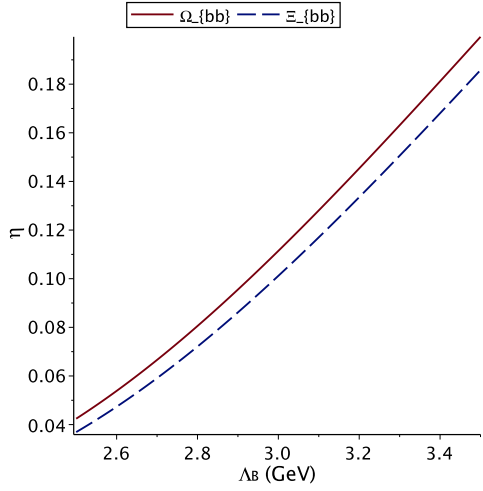}
\caption{ Behavior of $ \eta $ function versus $ \Lambda_B $ for $\Omega_{bb}\rightarrow \Omega_{bc}\ell \bar {\nu}_{\ell}$ and $\Xi_{bb}\rightarrow \Xi_{bc}\ell \bar {\nu}_{\ell}$ ($\ell=e$ or $\mu$) transitions ($ \omega=\omega_{\rm max} $). }
\label{fig:4}
\end{figure}

Using the obtained masses listed in Table~\ref{tab:mass1}
and Eqs.~(\ref{gamma11})-(\ref{l4}), the semileptonic decay
rates of doubly heavy  $ \Xi $ and $ \Omega $ baryons are
calculated.  The  $ \omega $ dependence of the
semileptonic decay widths
for $\Omega_{bb}\rightarrow \Omega_{bc}\ell \bar {\nu}_{\ell}$, $\Xi_{bb}\rightarrow \Xi_{bc}\ell \bar {\nu}_{\ell}$,
$\Omega_{bc}\rightarrow \Omega_{cc}\ell \bar {\nu}_{\ell}$, and $\Xi_{bc}\rightarrow \Xi_{cc}\ell \bar {\nu}_{\ell}$  transitions
is shown in Fig.~\ref{fig:2} and ~\ref{fig:3}, respectively, where we take $ \Lambda_B=3 $ GeV and
neglect the mass difference between
the $ u $ and $ d $  quarks. Regarding $\omega=\omega_{\rm max}$,  the $ \Lambda_B $
dependence of $ \eta $ functions for
 $\Omega_{bb} \rightarrow \Omega_{bc}$ and $\Xi_{bb}\rightarrow \Xi_{bc}$
transitions is shown in Fig.~\ref{fig:4}. Note that a smaller value
of $ \Lambda_B $ gives smaller decay rates and vice versa.

The calculated semileptonic decay widths of doubly heavy baryons
 and their variations are summarized in Table~\ref{tab:semileptonic1}.
The uncertainties in decay widths are due to the
parameter $\Lambda_{B} $, which can take a value in the range $ 2.5
\leq \Lambda_B \leq 3.5$ GeV. The uncertainties from the quark masses and potential parameters determined by fitting to the experimental data ~\cite{ghalenovi2018} can be neglected. A comparison between our results and those derived by Refs.~\cite{Hernandez2008,Faessler2009,Ebert2004,hassanabadi2020,Wang2017}
 is also presented.
For $ s=\frac{1}{2}\rightarrow s=\frac{1}{2} $ transitions,
including $ \Xi_{bb}\rightarrow \Xi_{bc} $, $\Xi_{bc}\rightarrow
\Xi_{bb} $, $ \Omega_{bb}\rightarrow \Omega_{bc} $, and
$ \Omega_{bc}\rightarrow \Omega_{cc} $, our results are
 in good agreement with those of Ref.~\cite{Faessler2009}.
In the case of $ \Xi_{bb}\rightarrow \Xi_{bc} $ and
$ \Omega_{bb}\rightarrow \Omega_{bc} $ transitions our
 results are  close to those of Ref.~\cite{Ebert2004}.

\begin{table} [H]
    \caption{ Semileptonic decay widths of doubly heavy baryons in units of $10^{-14}$ GeV and their variations.}
      \label{tab:semileptonic1}
    \centering
    \begin{center}
    \begin{tabular}{ccccccc}\hline

     Decay~~& ~Our  results~~& ~RCQM~\cite{Faessler2009}~&~NRQM~\cite{hassanabadi2020}~ & ~RQM~\cite{Ebert2004}~~ & ~HQSS~\cite{Hernandez2008}~& ~~LFQM\footnotemark[1]~\cite{Wang2017}~  \\
\hline
$\Xi_{bb}\rightarrow \Xi_{bc}\ell \bar {\nu}_{\ell}$ & ~~$1.75\pm 0.73 $~   & ~$1.33\pm 0.61 $&  1.37  & 1.63& 1.92$ ^{+0.25}_{-0.05} $&3.30\\
$\Xi_{bc}\rightarrow \Xi_{cc}\ell \bar {\nu}_{\ell}$ & ~~$ 4.39\pm0.83 $~    &  ~$4.01\pm 1.21 $& 5.07 & 2.30 &2.57$ ^{+0.26}_{-0.03} $&4.50\\
$\Xi_{bb}^*\rightarrow \Xi_{bc}\ell \bar {\nu}_{\ell}$ &  ~~$ 0.49\pm0.18 $~   &  ~$0.25\pm 0.10 $& 0.66  &0.28 & 0.35$^{+0.03}$&\\
$\Xi_{bc}^*\rightarrow \Xi_{cc}\ell \bar {\nu}_{\ell}$ & ~~$ 1.00\pm0.16 $~  &  ~$0.58\pm 0.14 $& 1.16 &0.38& 0.43$^{+0.06}$&\\
$\Xi_{bb}\rightarrow \Xi_{bc}^*\ell \bar {\nu}_{\ell}$ &  ~~$ 1.07\pm0.43 $~  & ~$0.61\pm 0.15 $& 1.45 &0.53&  0.61$^{+0.04}$&\\
$\Xi_{bc}\rightarrow \Xi_{cc}^*\ell \bar {\nu}_{\ell}$ &  ~~$ 2.82\pm0.52 $~  &   ~$1.39\pm 0.34 $& 3.32 &0.72& 0.75$^{+0.06}$&\\
$\Xi_{bb}^*\rightarrow \Xi_{bc}^*\ell \bar {\nu}_{\ell}$ & ~~$ 1.10\pm0.48 $~ &    ~$1.62\pm 0.73 $&   &1.92& 2.09$^{+0.16}$&\\
$\Xi_{bc}^*\rightarrow \Xi_{cc}^*\ell \bar {\nu}_{\ell}$ & ~~$ 3.12\pm0.62 $~  &    ~$4.63\pm 1.23 $&   & 2.69&2.63$^{+0.40}$&\\
\hline
$\Omega_{bb}\rightarrow \Omega_{bc}\ell \bar {\nu}_{\ell}$ &  ~~$ 1.87\pm 0.76 $ ~  &~$1.92\pm 1.15 $ &2.48   & 1.70&2.14$ ^{+0.20}_{-0.02} $ &3.69\\
$\Omega_{bc}\rightarrow \Omega_{cc}\ell \bar {\nu}_{\ell}$ &  ~~$4.70\pm0.83$~  &  ~$4.12\pm 1.10 $ & 5.39  & 2.48&2.59$ ^{+0.20} $ &3.94\\
$\Omega_{bb}^*\rightarrow \Omega_{bc}\ell \bar {\nu}_{\ell}$ & ~~$ 0.53\pm0.19 $ ~   & ~$0.26\pm 0.10 $ &0.69 & 0.29   &0.38$ ^{+0.04}_{-0.02} $ & \\
$\Omega_{bc}^*\rightarrow \Omega_{cc}\ell \bar {\nu}_{\ell}$ & ~~$ 1.09\pm0.16 $~  & ~$0.59\pm 0.13 $& 1.23  &  0.40  &0.44$ ^{+0.06} $ & \\
$\Omega_{bb}\rightarrow \Omega_{bc}^*\ell \bar {\nu}_{\ell}$ &  ~~$ 1.14\pm0.45 $~  &  ~$0.57\pm 0.23 $& 1.53 &  0.55  &0.67$ ^{+0.08}$  &\\
$\Omega_{bc}\rightarrow \Omega_{cc}^*\ell \bar {\nu}_{\ell}$ &  ~~$ 2.99\pm0.52 $ ~&  ~$1.78\pm 0.64 $  & 3.52 & 0.74   &0.76$ ^{+0.13}$  &\\
$\Omega_{bb}^*\rightarrow \Omega_{bc}^*\ell \bar {\nu}_{\ell}$ & ~~$ 1.17\pm0.50 $~  &    ~$1.72\pm 0.77 $&  & 2.00   &2.29$ ^{+0.31}_{-0.04} $  &\\
$\Omega_{bc}^*\rightarrow \Omega_{cc}^*\ell \bar {\nu}_{\ell}$ & ~~$ 3.32\pm0.62 $ ~&   ~$4.95\pm 1.26 $&  &  2.88  &2.79$ ^{+0.60} $  &\\
\hline
    \end{tabular}
    \end{center}
    \footnotetext[1]{LFQM denotes the light-front quark model.}
\end{table}

The absolute branching fractions of semileptonic decays of doubly heavy
 baryons can be easily derived  by using the following relation
\begin{equation}
{\cal B}=\Gamma\times\tau,
\end{equation}
where $ \tau $ is the lifetime of the initial baryon. Taking $ \tau_{\Xi_{bb}}=370\times 10^{-15}s$, $  \tau_{\Xi_{bc}}=244 \times 10^{-15}s $~\cite{karliner2014}, $  \tau_{\Omega_{bc}}=220 \times 10^{-15}s $, and $  \tau_{\Omega_{bb}}=800 \times 10^{-15}s $~\cite{Kiselev2002,Kiselev2002b} as input values,
the calculated branching fractions of semileptonic decays of doubly heavy
 baryons are summarized in Table~\ref{Beanching}.

\begin{table}[H]
    \caption{The calculated branching fractions of semileptonic decays of doubly heavy baryons.}
      \label{Beanching}
    \centering
    \begin{center}
    \begin{tabular}{cccc} \hline
     Process~~& Our results~~& NRQM~\cite{hassanabadi2020}~~ & LFQM~\cite{Wang2017}~~ \\
\hline
$\Xi_{bc}\rightarrow \Xi_{cc}\ell \bar {\nu}_{\ell}$~ & $1.63\times10^{-2}$ &$1.11\times10^{-2}$&$1.67\times10^{-2}$\\
$\Xi_{bb}\rightarrow \Xi_{bc}\ell \bar {\nu}_{\ell}$~ & $0.98\times10^{-2}$ &$0.28\times10^{-2}$&$1.86\times10^{-2}$\\
$\Omega_{bc}\rightarrow \Omega_{cc}\ell \bar {\nu}_{\ell}$~ & $1.57\times10^{-2}$ &$1.10\times10^{-2}$&$1.32\times10^{-2}$\\
$\Omega_{bb}\rightarrow \Omega_{bc}\ell \bar {\nu}_{\ell}$~ &$2.27\times10^{-2}$&&$4.49\times10^{-2}$\\
\hline
    \end{tabular}
    \end{center}
\end{table}

\section{Summary} \label{Summary}
In this paper we present a phenomenological
 study of the mass spectra and semileptonic decays of
 doubly heavy $\Xi$ and $\Omega$ baryons. The three-body problem of
these baryons is considered in the hypercentral approach.
 Applying an Ansatz method introduced in our previous
 work, the mass spectra of the ground and excited
 states of doubly heavy $\Xi$ and $\Omega$ baryons are obtained. By introducing a
 simple form for the universal IW function, we also
investigate the semileptonic decay rates and branching
fractions of the $ bb\rightarrow bc $ and $ bc\rightarrow cc $
 baryonic transitions near to zero recoil point.
We present a comparison between our results
and other available theoretical calculations, and find that the results
 are acceptable. Note that the triplet doubly heavy
baryons with $ s=\frac{3}{2} $ are dominated by the
strong or electromagnetic decays. If these excited
states are the initial ones, due to the smallness of
the weak coupling, the weak decays can not be observed
in the experiments. Therefore, one can shelve the
calculations for the semileptonic decays of
 $ s=\frac{3}{2} $ doubly heavy baryons, since
it is hard to perform relevant measurements  in the experiments. In the
current work, we have performed the related calculations
for all the ground states of doubly heavy baryons as done in Refs.~\cite{Faessler2009,Ebert2004,Hernandez2008}.
We hope our results are useful to extract the
value of the CKM matrix
element $ V_{cb} $ from future experiments via
the semileptonic decays of doubly heavy baryons. \\

\section{Acknowledgments}

The authors would like to thank Prof. Fu-Sheng Yu for
 useful discussions.
This work is supported in part by National Natural Science
Foundation of China (NSFC) under contract
No.~11975076, No.~12135005, and No.~12161141008.


\begin{thebibliography}{**}

\bibitem{Weng2018}
X. Z. Weng, X. L. Chen, and W. Z. Deng, Phys. Rev. D \textbf{97}, 054008 (2018).

\bibitem{Garcilazo2016}
H. Garcilazo, A. Valcarce, and J. Vijande, Phys. Rev. D \textbf{94}, 074003 (2016).

\bibitem{Shaha2017}
Z. Shah and A. K. Rai, Eur. Phys. J. C \textbf{77}, 129 (2017).

\bibitem{Brown2014}
Z. S. Brown, W. Detmold, S. Meinel, and K. Orginos, Phys. Rev. D \textbf{90}, 094507 (2014).

\bibitem{Aliev12013}
T. M. Aliev, K. Azizi, and M. Savci, J. Phys. G: Nucl. Part. Phys. \textbf{40}, 065003 (2013).

\bibitem{Mattson2002}
M. Mattson {\it et al.} (SELEX Collaboration), Phys. Rev. Lett. \textbf{89}, 112001 (2002).

\bibitem{Ocherashvili2005}
A. Ocherashvili {\it et al.} (SELEX Collaboration), Phys. Lett. B \textbf{628}, 18 (2005).

\bibitem{nfocus} S. P. Ratti, Nucl. Phys. Proc. Suppl. {\bf 115}, 33 (2003).

\bibitem{nbabar} B. Aubert {\it et al.} (BaBar Collaboration), Phys. Rev. D {\bf 74}, 011103 (2006).

\bibitem{nbelle} Y. Kato {\it et al.} (Belle Collaboration), Phys. Rev. D {\bf 89}, 052003 (2014).

\bibitem{LHCb2017}
R. Aaij {\it et al.} (LHCb Collaboration), Phys. Rev. Lett. \textbf{119}, 112001 (2017).

\bibitem{LHCb2018}
R. Aaij {\it et al.} (LHCb Collaboration), Phys. Rev. Lett. \textbf{121}, 162002 (2018).

\bibitem{LHCb20188}
R. Aaij {\it et al.} (LHCb Collaboration), Phys. Rev. Lett. \textbf{121}, 052002 (2018).

\bibitem{lhcb1} R. Aaij {\it et al.} (LHCb Collaboration), J. High Energy Phys. {\bf 11}, 095 (2020).


\bibitem{lhcb2} R. Aaij {\it et al.} (LHCb Collaboration), Chin. Phys. C {\bf 45}, 093002 (2021).

\bibitem{qqin} Q. Qin, Y. J. Shi, W. Wang, G. H. Yang, F. S. Yu, and R. L. Zhu, Phys. Rev. D {\bf 105}, L031902
(2022).

\bibitem{Li2021}
Y. S. Li, X. Liu, and F. S. Yu, Phys. Rev. D \textbf{104}, 013005 (2021).

\bibitem{Hsiao2020}
Y. K. Hsiao, L. Yang, C. C. Lih, and S. Y. Tsai, Eur. Phys. J. C \textbf{80}, 1066 (2020).

\bibitem{Lu2021}
C. D. L\"u, W. Wang, and F. S. Yu, Phys. Rev. D \textbf{93}, 056008 (2016).

\bibitem{Faustov2019}
R. N. Faustov and V. O. Galkin, Eur. Phys. J. C \textbf{79}, 695 (2019).

\bibitem{Faustov2016}
R. N. Faustov and V. O. Galkin, Eur. Phys. J. C \textbf{76}, 628 (2016).

\bibitem{Gutsche2016}
T. Gutsche, M. A. Ivanov, J. G. K\"orner, V. E. Lyubovitskij, and P. Santorelli, Phys. Rev. D \textbf{90}, 114033 (2014) [Erratum: Phys. Rev. D {\bf 94}, 059902 (2016)];
Phys. Rev. D \textbf{93}, 034008 (2016).

\bibitem{Ablikim2017}
M. Ablikim {\it et al.}  (BESIII Collaboration), Phys. Lett. B \textbf{767}, 42 (2017).

\bibitem{Ablikim2018}
M. Ablikim {\it et al.}  (BESIII Collaboration), Phys. Rev. Lett. \textbf{121}, 251801 (2018).

\bibitem{Li2021a}
Y. B. Li {\it et al.}  (Belle Collaboration), Phys. Rev. Lett. \textbf{127}, 121803 (2021).

\bibitem{Li2021b}
Y. B. Li {\it et al.}  (Belle Collaboration), Phys. Rev. D {\bf 105}, L091101 (2022).

\bibitem{Ammar2002}
R. Ammar {\it et al.}  (CLEO Collaboration), Phys. Rev. Lett. \textbf{89}, 171803 (2002).

\bibitem{Kiselev2001}
V. V. Kiselev and A. E. Kovalsky, Phys. Rev. D \textbf{64}, 014002 (2001).

\bibitem{Shi2020b}
Y. J. Shi, W. Wang, Z. X. Zhao, and U. G. Mei\ss{}ner, Eur. Phys. J. C \textbf{80}, 398 (2020).

\bibitem{Gutsche2019}
T. Gutsche, M. A. Ivanov, J. G. K\"orner, V. E. Lyubovitskij, and Z. Tyulemissov, Phys. Rev. D \textbf{100}, 114037 (2019).


\bibitem{Yu2019}
  Q. X. Yu and X. H. Guo, Nucl. Phys. B \textbf{947}, 114727 (2019).

\bibitem{Hernandez2008}
E. Hernandez, J. Nieves, and J. M. Verde-Velasco, Phys. Lett. B \textbf{663}, 234 (2008).

\bibitem{Lyubovitskij2003}
V. E. Lyubovitskij, A. Faessler, T. Gutsche, M. A. Ivanov, and J. G. K\"orner, Prog. Part. Nucl. Phys. \textbf{50}, 329 (2003).

\bibitem{Lyubovitskij2001}
A. Faessler, T. Gutsche, M. A. Ivanov, J. G. K\"orner and V. E. Lyubovitskij, Phys. Lett. B \textbf{518}, 55 (2001).

\bibitem{Albertus2008}
 C. Albertus, E. Hernandez, J. Nieves, and J. M. Verde-Velasco, Eur. Phys. J. A \textbf{32}, 183 (2007),
  [Erratum]: Eur.Phys.J.A \textbf{36}, 119 (2008).


\bibitem{Shi2020a}
Y. J. Shi, W. Wang, Z. X. Zhao, and U. G. Mei\ss{}ner, Eur. Phys. J. C \textbf{80}, 398 (2020).


\bibitem{ghalenovi2018}
Z. Ghalenovi and M. M. Sorkhi, Eur. Phys. J. Plus \textbf{133}, 301 (2018).


\bibitem{Faessler2009}
A. Faessler, T. Gutsche, M. A. Ivanov, J. G. K\"orner, and V. E. Lyubovitskij, Phys. Rev. D \textbf{80}, 034025 (2009).


\bibitem{Isgur1978}
N. Isgur and G. Karl, Phys. Rev. D \textbf{18}, 4187 (1978).

\bibitem{Capstick1986}
S. Capstick and N. Isgur, Phys. Rev. D \textbf{34}, 2809 (1986).

\bibitem{Ferretti2011}
J. Ferretti, A. Vassallo, and E. Santopinto, Phys. Rev. C \textbf{83}, 065204 (2011).

\bibitem{Ghalenovi:2012iu}
  Z.~Ghalenovi and A.~A. Rajabi,
  Eur.\ Phys.\ J.\ Plus {\bf 127}, 141 (2012).

\bibitem{Lonsdale:2017mzg}
S.~J.~Lonsdale, M.~Schroor, and R.~R.~Volkas,
Phys. Rev. D \textbf{96}, 055027 (2017).

\bibitem{Menapara:2021dzi}
C.~Menapara and A.~K.~Rai,
Chin. Phys. C \textbf{45},  063108 (2021).

\bibitem{ansatz1} A. A. Rajabi, Few-Body Systems {\bf 37}, 197 (2005).

\bibitem {Ghalenovi:2014swa}
Z.~Ghalenovi, A.~A.~Rajabi, S.~X.~Qin, and D.~H.~Rischke,
Mod. Phys. Lett. A \textbf{29}, 1450106 (2014).

\bibitem{Ghalenovi:2017xxv}
Z.~Ghalenovi and M.~Moazzen,
Eur. Phys. J. Plus \textbf{132}, 354 (2017).

\bibitem{Ebert2004}
D. Ebert, R. N. Faustov, V. O. Galkin, and A. P. Martynenko, Phys. Rev. D \textbf{70}, 014018 (2004)
[Erratum: Phys. Rev. D {\bf 77}, 079903 (2008)].

\bibitem{Albertus2010}
C. Albertus, E. Hern{\'a}ndez, and J. Nieves, Phys. Lett. B \textbf{683}, 21 (2010).

\bibitem{Azizi2013}
T. M. Aliev, K. Azizi, and M. Savei, J. Phys. G \textbf{40}, 065003 (2013).

\bibitem{Roberts2008}
W. Roberts and M. Pervin, Int. J. Mod. Phys. A \textbf{23}, 2817 (2008).

\bibitem{Yoshida2015}
T. Yoshida, E. Hiyama, A. Hosaka, M. Oka, and K. Sadato, Phys. Rev. D \textbf{92}, 114029 (2015).

\bibitem{Shah2017}
Z. Shah and A. K. Rai, Eur. Phys. J. C \textbf{77}, 129  (2017).

\bibitem{Soto2021}
J. Soto and J. T. Castell\`a, Phys. Rev. D \textbf{104}, 074027 (2021).

\bibitem{Shah2016}
Z. Shah, K. Thakkar, and A. K. Rai, Eur. Phys. J. C \textbf{76}, 530  (2016).

\bibitem{Georgi1990}
H. Georgi and M. B. Wise, Phys. Lett. B \textbf{243}, 279 (1990).

\bibitem{Carone1991}
C. D. Carone, Phys. lett. B \textbf{253}, 408 (1991).

\bibitem{Flynn2008}
J. M. Flynn and J. Neives, Phys. Rev. D \textbf{76}, 017502 (2007) [Erratum: Phys. Rev. D \textbf{77}, 099901 (2008].

\bibitem{Isgur1991}
N. Isgur and M. B. Wise, Nucl. Phys. B \textbf{348}, 276 (1991).

\bibitem{Ebert20066}
D. Ebert, R. N. Faustov and V. O. Galkin, Phys. Rev. D \textbf{73}, 094002 (2006).

\bibitem{hassanabadi2020}
S. Rahmani, H. Hassanabadi, and  H. Sobhani, Eur. Phys. J. C \textbf{80}, 312 (2020).


\bibitem{Wang2017}
W. Wang, F. S. Yu, and Z. X. Zhao, Eur. Phys. J. C \textbf{77}, 781 (2017).

\bibitem{karliner2014}
M. Karliner and J. L. Rosner, Phys. Rev. D \textbf{90}, 094007 (2014).

\bibitem{Kiselev2002}
V. V. Kiselev and A. K. Likhoded, Phys. Usp.  \textbf{172}, 497 (2002).

\bibitem{Kiselev2002b}
V. V. Kiselev and A. K. Likhoded, Phys. Usp. \textbf{45}, 455 (2002).



\end{thebibliography}
\end{document}